# Accurate Measurement of the Electron Antineutrino Yield of $^{235}$U Fissions from the STEREO Experiment with 119 Days of Reactor-On Data

H. Almazán,[1,*] L. Bernard,[2,†] A. Blanchet,[3,‡] A. Bonhomme,[1,3] C. Buck,[1] P. del Amo Sanchez,[4] I. El Atmani,[3,§] J. Haser,[1] L. Labit,[4] J. Lamblin,[2] A. Letourneau,[3,∥] D. Lhuillier,[3] M. Licciardi,[2] M. Lindner,[1] T. Materna,[3] A. Minotti,[3,¶] A. Onillon,[3] H. Pessard,[4] J.-S. Réal,[2] C. Roca,[1] R. Rogly,[3] T. Salagnac,[2,**] V. Savu,[3] S. Schoppmann,[1,††] V. Sergeyeva,[4,‡‡] T. Soldner,[5] A. Stutz,[2] and M. Vialat[5]

(STEREO Collaboration)

[1]*Max-Planck-Institut für Kernphysik, Saupfercheckweg 1, 69117 Heidelberg, Germany*
[2]*Université Grenoble Alpes, CNRS, Grenoble INP, LPSC-IN2P3, 38000 Grenoble, France*
[3]*IRFU, CEA, Université Paris-Saclay, 91191 Gif-sur-Yvette, France*
[4]*Université Grenoble Alpes, Université Savoie Mont Blanc, CNRS/IN2P3, LAPP, 74000 Annecy, France*
[5]*Institut Laue-Langevin, CS 20156, 38042 Grenoble Cedex 9, France*

We report a measurement of the antineutrino rate from the fission of $^{235}$U with the STEREO detector using 119 days of reactor turned on. In our analysis, we perform several detailed corrections and achieve the most precise single measurement at reactors with highly enriched $^{235}$U fuel. We measure an IBD cross section per fission of $\sigma_f = (6.34 \pm 0.06[\text{stat}] \pm 0.15[\text{sys}] \pm 0.15[\text{model}]) \times 10^{-43}~\text{cm}^2/\text{fission}$ and observe a rate deficit of $(5.2 \pm 0.8[\text{stat}] \pm 2.3[\text{sys}] \pm 2.3[\text{model}])\%$ compared to the model, consistent with the deficit of the world average. Testing $^{235}$U as the sole source of the deficit, we find a tension between the results of lowly and highly enriched $^{235}$U fuel of 2.1 standard deviations.

DOI: 10.1103/PhysRevLett.125.201801

In recent years, neutrino physics at nuclear reactors has entered a precision era. The neutrino mixing angle $\theta_{13}$ was determined and constraints of the absolute antineutrino rate were achieved [1,2]. Experiments at reactors with highly and lowly enriched $^{235}$U fuel [2–5] confirm the ∼6% deficit of observed electron antineutrinos when compared to state-of-the-art antineutrino energy spectrum calculations, known as the reactor antineutrino anomaly (RAA) [6,7]. The anomaly has triggered numerous works to find explanations. The existence of a sterile neutrino state is explored by several short baseline experiments [8–11]. STEREO is one of them, searching for a nonstandard oscillation in the propagation of the electron antineutrino at ∼10 m baseline [12,13]. The Daya Bay and RENO collaborations have reported an observation of correlation between the reactor core evolution and changes in the deficit of the reactor antineutrino flux [3,4]. They conclude that $^{235}$U might be the primary contributor to the RAA. However, a contribution of $^{239}$Pu cannot be ruled out [14]. Updated antineutrino spectrum predictions argue for larger model uncertainties or yield a smaller deficit [15–17].

In this context, we report a precision measurement of the electron antineutrino yield with the STEREO experiment at a reactor using highly enriched $^{235}$U fuel as well as a comparison between the mean inverse $\beta$ decay (IBD) cross sections of $^{235}$U derived from experiments at reactors with highly and lowly enriched $^{235}$U fuel, respectively. The measurement is based on 119 days of reactor-on and 211 days of reactor-off data (STEREO phase II as defined in [13], Sec. III) with a high detector stability ([13], Sec. VI), providing 43 400 detected antineutrino events [18–21]. STEREO [22] is installed at the high flux reactor (RHF, Réacteur à Haut Flux [23,24]) of the Institut Laue-Langevin. The RHF operates with a $^{235}$U enrichment of 93%, thus 99.3% of the electron antineutrino flux is produced by fissions of $^{235}$U. STEREO is situated below a water-filled transfer channel that mitigates cosmic-induced radiations.

STEREO consists of a target volume filled with organic liquid scintillator loaded with gadolinium (Gd). It is surrounded by a gamma catcher filled with unloaded liquid scintillator. The target scintillator is composed of linear alkylbenzene (LAB), ortho-phenylxylylethane (PXE), and diisopropylnaphtalene (DIN) [25]. It acts as a proton reservoir to detect electron antineutrinos via the IBD reaction on hydrogen nuclei: $\bar{\nu}_e + p \rightarrow e^+ + n$. The target volume is divided into six identical and optically separated cells. In each cell, light pulses are recorded by four 8-in. photomultiplier tubes mounted above a 20-cm-thick acrylic buffer. In the following, we detail the calculation of the expectation of the antineutrino rate and describe its measurement.

Since, in a nuclear reactor, electron antineutrinos are produced by $\beta^-$ decays of fission fragments in the reactor core, their total number over one cycle can be written in good approximation as

$$N_\nu^{\text{emi}} = \frac{\langle P_{\text{th}} \rangle}{\langle E_f \rangle} \iint \sum_i [f_i(t) S_i(E_\nu)] dE_\nu dt \times (1 + c_{\text{SNF}}), \quad (1)$$

where $\langle P_{\text{th}} \rangle$ is the mean reactor thermal power from nuclear reactions, $\langle E_f \rangle$ is the mean energy released per fission, $f_i(t)$ is the activity per fission of the $i$th $\beta$ emitter, $S_i(E_\nu)$ is the associated antineutrino energy spectrum, and $c_{\text{SNF}}$ is a correction due to the contribution of the spent nuclear fuel.

The first term in Eq. (1) expresses the number of fissions. It is based on the assumption that all the energy produced in one fission is converted into heat in the installation. It is measured integrally. By simulating the RHF in high detail using the MCNPX-2.5 [26] and TRIPOLI-4 [27] codes, we find the amount of energy loss by escaping neutrons and $\gamma$ rays negligible. We can thus use the total thermal power $P_{\text{th}}^{\text{tot}}$ measured by the RHF and subtract the mechanical power of the water flow $P_{\text{pumps}}$, which dissipates inside the moderator tank,

$$P_{\text{th}} = P_{\text{th}}^{\text{tot}} - P_{\text{pumps}}, \quad (2)$$

with $P_{\text{pumps}} = (0.7 \pm 0.1)$ MW [28]. The computation of the total thermal power is based on the general equation

$$P_{\text{th}}^{\text{tot}} = \sum_c \{q_v \times [\rho(T_d) C_p(T_d) T_d - \rho(T_u) C_p(T_u) T_u]\}, \quad (3)$$

where $q_v$ is the volumic flow rate, $\rho$ is the volumic density of the water, $C_p$ is the calorific capacity, and $T$ is the temperature. The indices $u$ and $d$ denote quantities measured upstream and downstream of the moderator tank, respectively. The sum runs over four instrumented circuits $c$ of fluids, of which the primary heavy water circuit carries 96% of the total power. The main flow rate measurement is based on the Venturi effect induced by a calibrated diaphragm inserted in the primary circuit. All temperature and pressure sensors are duplicated for cross monitoring and they are accurately calibrated every two years. Propagating all uncertainties leads to a 1.44% relative accuracy [29] with a mean power during reactor-on periods used in this analysis of $\langle P_{\text{th}}^{\text{tot}} \rangle = (49.9 \pm 0.7)$ MW. A significant contribution to the total relative uncertainty (0.9%) comes from the calibration of the diaphragm, performed in the 1970s with a scale 1 mock-up of the primary circuit [30]. Since diaphragms at power reactors did not show any aging effects [31], we assume that the accuracy of this calibration still holds. The possibility of a cross-check by an inspection of the diaphragm during a reactor shutdown is under investigation.

The mean energy released per fission $\langle E_f \rangle$ is a key ingredient to extract the number of fissions from the measured thermal power. Precise values were obtained by Ma et al. [32] using the mass conservation method proposed by Kopeikin et al. [33], where the energy release per fission is written as

$$E_f = E_{\text{tot}} - \langle E_\nu \rangle - \Delta E_{\beta\gamma} + E_{\text{nc}} \quad (4)$$

and is based on $E_{\text{tot}}$, the mass excess difference between the initial and the final fragmented systems after all fragments have decayed. Corrections are applied to take into account the energy loss by antineutrinos $\langle E_\nu \rangle$, the fraction of energy not released in the reactor due to long-lived fragments $\Delta E_{\beta\gamma}$, and the energy added due to radiative neutron captures on structural elements $E_{\text{nc}}$. All the terms depend on the irradiation conditions. In the work of Ma et al., the two latter terms were evaluated for a fuel irradiation time corresponding to the midpoint of a standard cycle of a pressurized water reactor [about 1.5 yr cycle duration giving $\Delta E_{\beta\gamma} = (0.35 \pm 0.02)$ MeV for $^{235}$U] and for a wide range of reactor materials [$E_{\text{nc}} = (8.57 \pm 0.22)$ MeV]. We recalculated these two values for our experimental conditions of irradiation period (50 days) and of dominance of aluminium as structural material in the core and moderator tank. Using recent databases (JEFF-3.3 [34], GEFY-6.2 [35], and NUBASE2016 [36]) and a precise TRIPOLI-4 simulation of the RHF (to model the activation of structural materials) ([13], Sec. IV), we evaluated these quantities to be $\Delta E_{\beta\gamma} = (0.6 \pm 0.1)$ and $E_{\text{nc}} = (10.3 \pm 0.2)$ MeV. The recent nuclear databases were also used to calculate an updated mass excess for the fission products of $^{235}$U. The obtained mean value using the cumulative fission yields from JEFF-3.3 and GEFY-6.2 amounts to $(-173.15 \pm 0.07)$ MeV. This value has to be compared with $(-173.86 \pm 0.06)$ MeV from the work of Ma et al., using the cumulative fission yields from JEFF-3.1 [37] and mass excesses from AME2003 [38] nuclear databases. This difference has to be considered as a bias on the value from the work of Ma et al. We note that the energy loss by antineutrinos requires extrapolations to energies below 2 MeV. In that region, the accumulation of long-lived isotopes produced by the $\beta$ decay of fission fragments or neutron captures modifies the antineutrino energy spectrum compared to the instantaneous one. For that reason, the extrapolation using exponential functions fitted on the energy spectrum above 2 MeV and measured after a few hours, as done in the method used by Ma et al., may not be a good estimate for the shape. A full simulation with all $\beta$ decays involved in the reactor core assuming a correct modeling of the shapes of all $\beta$ branches is required, but is unreliable at present. In the near future, progress in the summation method may refine the evaluation done by Ma

PHYSICAL REVIEW LETTERS **125,** 201801 (2020)

*et al.* The relative distortion of the antineutrino energy spectrum as a function of time, due to accumulation of long-lived isotopes and transmutations by neutron captures, were calculated with the FISPACT-II code coupled to the BESTIOLE code [39]. The averaged correction over one cycle amounts to 490 keV and we use a value of $\langle E_\nu \rangle = (9.55 \pm 0.13)$ MeV for $^{235}$U, the uncertainty covering the different reactor operations. In the following, and to be compatible with the previous work, we used the values from Ma *et al.* except for $\langle E_\nu \rangle$, $\Delta E_{\beta\gamma}$, and $E_{\mathrm{nc}}$, which are specific to our irradiation conditions. Likewise, the corresponding values for $^{239}$Pu were updated. The contribution of $^{239}$Pu was calculated using the FISPACT-II evolution code [40]. It was found to be 1.4% by the end of a nominal cycle, resulting in a mean contribution of only 0.7% [13]. By using this weighting, the mean energy released per fission amounts to $\langle E_f \rangle = (203.41 \pm 0.26)$ MeV.

The Huber spectrum for pure $^{235}$U [41] is used as a model of the integral in Eq. (1). As the Huber model is defined in the [2, 8] MeV range, we restrict our analysis to this energy range. To allow a better comparison between experiments, we do not correct this model from the distortions in the energy spectrum seen by several experiments [5,42]. The model is, however, corrected for the fission fraction of $^{239}$Pu, the time evolution of fission fragment activities, and activation of structural elements [13]. The fraction of $^{239}$Pu reduces the averaged antineutrino rate over one cycle by less than 0.3%. For the energy range of our analysis, it is found to affect mainly the three lowest 500-keV-wide energy bins above 2.4 MeV antineutrino energy. The maximum contribution of less than 2% is found in the first bin. The activation of structural materials was inferred using the TRIPOLI-4 simulation of the RHF. It was found that mainly $^{28}$Al and $^{56}$Mn contribute. Combining all low energy corrections leads to a sizeable increase of the total rate of emitted antineutrinos by $(7.2 \pm 0.4)\%$ with respect to the Huber model [13,43]. Because of the lower IBD cross section at low antineutrino energy, the impact on the predicted number of detected antineutrinos per fission is smaller, about $(1.6 \pm 0.1)\%$ neglecting experimental thresholds and cut efficiencies. The extra uncertainty is negligible compared to the initial uncertainty of 2.4% of the Huber model (see Table I). Finally, $c_{\mathrm{SNF}}$ in Eq. (1) arises from spent fuel elements stored in the transfer channel above the STEREO detector. Its relative contribution to the number of emitted neutrinos from the core was estimated with FISPACT-II coupled to BESTIOLE to be less than 0.1% after 24 h of a reactor stop, justifying that in our analysis only data after this time are considered. The remaining effect is further suppressed in the analysis by a factor $s_{\mathrm{SNF}}^{\mathrm{on\text{-}off}}$ due to the subtraction of reactor-on and -off data. Caused by, e.g., displacement of the stored fuel elements and time evolution, the residual effect amounts to less than 0.2% in the lowest 500-keV-wide energy bin [43].

TABLE I. Summary of all relevant quantities and their corresponding relative uncertainties on the IBD yield.

| Quantity | Symbol | Value | Uncertainty (%) |
|---|---|---|---|
| Number of $\nu$/fission | $N_\nu^{[2,8]\,\mathrm{MeV}}$ | 1.846 | 2.40 |
|    Huber prediction | | 1.722 | 2.40 |
|    Correction factors | | 1.072 | 0.10 |
| Number of fissions/day | | $1.30 \times 10^{23}$ | 1.44 |
|    Thermal power | $\langle P_{\mathrm{th}} \rangle$ | 49.2 MW | 1.44 |
|    Energy per fission | $\langle E_f \rangle$ | 203.4 MeV | 0.13 |
| Fraction of interacting $\nu$ | $\tau_{\mathrm{int}}$ | $8.10 \times 10^{-21}$ | 0.56 |
|    Solid angle | | | 0.50 |
|    IBD cross section | $\sigma_{\mathrm{IBD}}$ | | 0.22 |
|    MC statistics | | | 0.12 |
| Correction of $p$ number | $c_p^{\mathrm{Data/MC}}$ | 0.983 | 1.00 |
| Detection efficiency | $\epsilon_d$ | 0.2049 | 0.54 |
|    Selection cuts | | | 0.41 |
|    Energy scale | | | 0.30 |
|    MC statistics | | | 0.19 |
| Correction of delayed efficiency | $c_n^{\mathrm{Data/MC}}$ | 0.9774 | 0.86 |
| Predicted IBD yield | | 383.7 d$^{-1}$ | 2.10 ⊕ 2.40 |
| Observed IBD yield | | 363.8 d$^{-1}$ | 0.88 ⊕ 1.06 |
|    Statistics | | | 0.88 |
|    $\nu$ extraction method | | | 0.65 |
|    Reactor-induced background | | | 0.83 |
|    Off-time method | | | 0.14 |

From the total number of emitted antineutrinos, the predicted number of detected antineutrinos can be written as

$$N_\nu^{\text{pred}} = N_\nu^{\text{emi}} \times \tau_{\text{int}} \times s_{\text{SNF}}^{\text{on-off}} \times c_p^{\text{Data/MC}} \times \epsilon_d \times c_n^{\text{Data/MC}} \quad (5)$$

with the suppression factor $s_{\text{SNF}}^{\text{on-off}}$ as described above, as well as the fraction of interacting antineutrinos $\tau_{\text{int}}$, the proton number correction $c_p^{\text{Data/MC}}$, the total detection efficiency $\epsilon_d$, and the correction of the detection efficiency of the delayed signal $c_n^{\text{Data/MC}}$. These quantities are tabulated in Table I and discussed in the following.

The fraction of antineutrinos which interact in the detector can be written as

$$\tau_{\text{int}} = \iiint S(E_\nu)\sigma_{\text{IBD}}(E_\nu)\frac{\rho_f(\vec{r}_c)\rho_H(\vec{r}_d)}{4\pi||\vec{r}_d - \vec{r}_c||^2} d\vec{r}_d d\vec{r}_c dE_\nu, \quad (6)$$

where $S(E_\nu)$ is the antineutrino energy spectrum normalized by integral to unity, $\sigma_{\text{IBD}}(E_\nu)$ is the IBD cross section [44], $\vec{r}_c$ and $\vec{r}_d$ are the coordinates of the antineutrino emission and interaction vertices, $\rho_f(\vec{r}_c)$ is the fission density distribution in the core, normalized to unity and inferred from the MCNPX-2.5 simulation, and $\rho_H(\vec{r}_d)$ is the hydrogen density in the fiducial volume of the detector. This integral is numerically computed using a Monte Carlo (MC) method including the description of the reactor and detector setups. The emission vertices are generated randomly within the core following the fission density distribution. Likewise, also the interaction vertices are generated randomly within a portion of a hollow sphere enclosing the STEREO detector and following a $1/||\vec{r}_d - \vec{r}_c||^2$ distribution. The fraction of interactions $\tau_{\text{int}}$ has been found to be $(8.10 \pm 0.05) \times 10^{-21}$. The uncertainty on $\tau_{\text{int}}$ includes the geometrical solid angle uncertainty (0.50%), the IBD cross section uncertainty (0.22%), and a statistical uncertainty (0.12%) of the MC method [22]. It does not include the Huber model uncertainty. The factor $c_p^{\text{Data/MC}} = (0.983 \pm 0.010)$ corrects the number of hydrogen atoms used in the MC model to the one measured during detector filling [13,25].

In the experiment, IBD candidates are identified as two successive events within a time coincidence window of [2, 70] $\mu$s, passing energy cuts. These energy cuts are set to select the positron candidate (prompt signal) in the [1.625, 7.125] MeV energy range and the neutron candidate (delayed signal) in the [4.5, 10.0] MeV energy range ([13], Sec. VII). In addition to the basic selection cuts, a muon veto and topological selections are used to improve the accidental and correlated background rejections. All these rejection cuts induce detection inefficiencies that are calculated and propagated into the prediction ([13], Sec. VII).

The total detection efficiency $\epsilon_d$ is computed using GEANT4 [45,46] and FIFRELIN [47–49] simulations, as well as the same antineutrino generator as for the estimation of the fraction of interacting antineutrinos ([13], Sec. VII). This term describes for the antineutrinos of [2, 8] MeV kinetic energy, which interact in the scintillator or acrylics, the fraction passing all selection cuts. It accounts for detector effects such as energy nonlinearities or the energy resolution (energy dependent, better than 7% above 1.6 MeV [13], Sec. VI-C). It also evaluates the global efficiency of the prompt and delayed signals due to selection cuts, the fraction of neutron captures by Gd compared to other nuclei (mainly hydrogen) in the target, and the amount of events (either neutrons or $\gamma$ rays from the capture process) escaping to other volumes free of Gd. For the distribution of vertices obtained in the MC simulations, the total detection efficiency amounts to $\epsilon_d = (0.2049 \pm 0.0011)$. Even if the MC simulation has been extensively checked and tuned with a variety of calibration $\gamma$ sources, cell-dependent corrections to the neutron detection efficiency still have to be applied to correctly reproduce the neutron physics inside the detector. These corrections were evaluated using an AmBe $\gamma$-neutron source in the experiment and in the simulation ([13], Sec. VIII). The average correction factor of the detector between data and the MC simulation for the delayed signal amounts to $c_n^{\text{Data/MC}} = (0.9774 \pm 0.0084)$ ([50], Sec. 6.5).

Finally, the predicted antineutrino rate yields $(383.7 \pm 8.1[\text{sys}] \pm 9.2[\text{model}])\bar{\nu}_e/\text{day}$, where the experimental and Huber model uncertainties were separated, as the latter is common to all experiments. The experimental antineutrino rate is extracted by discriminating events via the pulse shape of the scintillation light of the prompt signal (PSD [13], Sec. IX-B). In this technique, we simultaneously fit the PSD distributions of IBD candidates measured during reactor-on and reactor-off periods ([13], Sec. X). Integrated over 119 days of reactor-on periods and 211 days of reactor-off periods, the IBD rate amounts to $(363.8 \pm 5.0)\bar{\nu}_e/\text{day}$. The uncertainty is due to statistics (0.88%), an added systematic uncertainty including systematic effects in the PSD fit, and covering small discrepancies when extracting the IBD rate with the method described in [51] (0.65%, corresponding to half of the discrepancy), another systematic uncertainty to cover the contribution of a possible reactor-induced background (0.83%) ([13], Sec. IX-C), and a systematic uncertainty to cover any potential bias in the off-time extraction method of accidental coincidences (0.14%) ([13], Sec. IX-A).

The comparison with the prediction gives an observed to predicted ratio of $0.948 \pm 0.008[\text{stat}] \pm 0.023[\text{sys}] \pm 0.023[\text{model}]$, where the first uncertainty is statistical, the second combines all experimental systematic uncertainties listed in Table I, and the third uncertainty is from the Huber model, common to all experiments. All systematic uncertainties are treated uncorrelated. Considering only the two experimental uncertainties, we find very good agreement with the world average of pure $^{235}$U measurements [2]. Our measurement is consistent with the deviation from the Huber model as shown in Fig. 1. Including our measurement, the world average is improved from $(0.950 \pm 0.015)$ to $(0.950 \pm 0.013)$, where again only experimental uncertainties are considered.

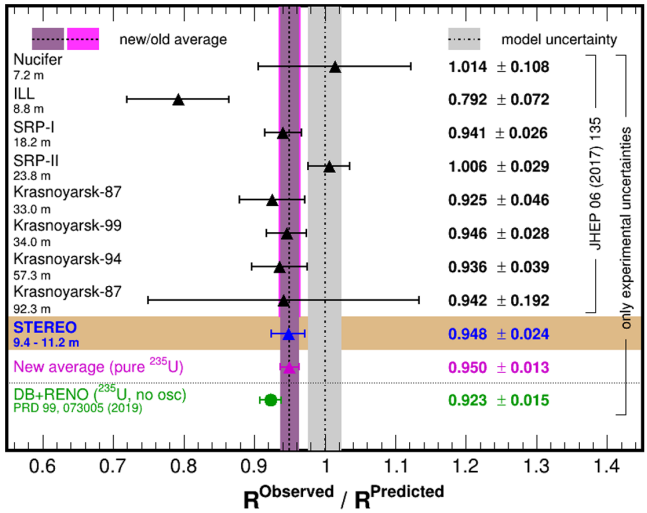

FIG. 1. Ratios between measured antineutrino yields and the Huber model predictions of various experiments. The uncertainty bars represent only experimental uncertainties. The common model uncertainty of 2.4% is shown as gray band around unity. Values of other experiments are taken from Refs. [2,14] (and references therein). For Daya Bay and RENO, we show only the ratio for the $^{235}$U component. The value is taken from a fit, where isotopic IBD yields of $^{235}$U and $^{239}$Pu are free, while those of $^{238}$U and $^{241}$Pu are constrained to the prediction [14].

To determine the IBD cross section per fission, we use an extrapolated Huber spectrum $S_H(E_\nu)$ for pure $^{235}$U without corrections for $^{28}$Al, $^{56}$Mn, and off-equilibrium effects. For the extrapolation, two independent exponential functions are fitted in the [2.0, 4.0] and [7.3, 8.0] MeV range, respectively. We find uncertainties related to the extrapolation negligible. The corresponding integral

$$\sigma_f^{\text{pred}} = \int_{1.8 \text{ MeV}}^{10.0 \text{ MeV}} S_H(E_\nu) \sigma_{\text{IBD}}(E_\nu) dE_\nu \qquad (7)$$

yields a predicted theoretical value fully consistent with the value of $(6.69 \pm 0.15) \times 10^{-43}$ cm$^2$/fission from Ref. [52], which we use in the following. Applying our observed to predicted ratio, we get $\sigma_f = (6.34 \pm 0.06[\text{stat}] \pm 0.15[\text{sys}] \pm 0.15[\text{model}]) \times 10^{-43}$ cm$^2$/fission, consistent with the value in [14].

Analyses of the dependence of the deficit on the core composition of power reactors [14] have indicated that $^{235}$U may be the main contributor responsible for the RAA. We extract $\sigma_{f,235}$ from measurements of $\sigma_f$ at power reactors [3–5,53] as

$$\sigma_{f,235} = \frac{\sigma_f - \sum_{x \in \{238,239,241\}}(\sigma_{f,x}^{\text{pred}} \alpha_x)}{\alpha_{235}}, \qquad (8)$$

assuming that other isotopes are within their predictions $\sigma_{f,x}^{\text{pred}}$ [52,54]. Taking into account experimental and model uncertainties, as well as correlations between the $\sigma_{f,x}^{\text{pred}}$ [54] and the constraint on the sum of all fission fractions $\alpha_x$ [3–5,53], we find a discrepancy of 2.1 standard deviations between the average of the power reactors and the new world average of highly enriched $^{235}$U reactors. This indicates that $^{235}$U may not be the only isotope responsible for the RAA.

The result presented in this Letter demonstrates the ability of STEREO to achieve an accurate measurement of the electron antineutrino rate coming from a pure $^{235}$U fuel element. It is consistent with the observed deficit from the Huber model corresponding to the RAA and is in agreement with the measured world average. While our result is already the most precise among all pure $^{235}$U measurements, some improvement is possible as additional data taking is in progress. Until the end of 2020, a twofold increase of the dataset is expected, allowing us to reduce statistical uncertainties and improve the sensitivity of our systematic studies.

We thank the ILL divisions DRe and DPT for their help in the precise determination of the reactor power and solid angle. Moreover, we are grateful for the technical and administrative support of the ILL for the installation and operation of the STEREO detector. This work is funded by the French National Research Agency (ANR) within the Project No. ANR-13-BS05-0007 and the "Investments for the Future" programs P2IO LabEx (ANR-10-LABX-0038) and ENIGMASS LabEx (ANR-11-LABX-0012). We further acknowledge the support of the CEA, in particular, the financial support of the Cross-Disciplinary Program on Numerical Simulation of CEA, the CNRS/IN2P3, and the Max Planck Society.

[*]Present address: Donostia International Physics Center, Paseo Manuel Lardizabal, 4, 20018 Donostia-San Sebastian, Spain.
[†]Present address: Ecole Polytechnique, CNRS/IN2P3, Laboratoire Leprince-Ringuet, 91128 Palaiseau, France.
[‡]Present address: LPNHE, Sorbonne Université, Université de Paris, CNRS/IN2P3, 75005 Paris, France.
[§]Present address: Hassan II University, Faculty of Sciences, Aïn Chock, BP 5366 Maarif, Casablanca 20100, Morocco.
[‖]alain.letourneau@cea.fr
[¶]Present address: Dipartimento di Fisica e Scienze della Terra, University of Ferrara and INFN, 44122 Ferrara, Italy.
[**]Present address: Institut de Physique Nucléaire de Lyon, CNRS/IN2P3, Université Lyon, Université Lyon 1, 69622 Villeurbanne, France.
[††]stefan.schoppmann@mpi-hd.mpg.de